\documentclass[aps,prl,twocolumn,showpacs]{revtex4}
\usepackage{graphicx}%
\usepackage{dcolumn}
\usepackage{amsmath}
\usepackage{amssymb}
\usepackage{bbold}
\usepackage{latexsym}
\usepackage[colorlinks,citecolor={blue},linkcolor={blue}]{hyperref}
\newcommand{\dg}{\dagger}
\newcommand{\up}{\uparrow}
\newcommand{\dwn}{\downarrow}

\newcommand{\la}{\langle}
\newcommand{\ra}{\rangle}
\begin{document}
\bibliographystyle{apsrev} 
\title{Mesoscopic effects in adiabatic spin pumping} 
\author{Prashant Sharma} 
\author{Piet W. Brouwer}
\affiliation{Laboratory of Atomic and Solid State Physics, Cornell
University, Ithaca, New York 14853-2501}
\date{\today}
\begin{abstract}
  We show that temporal shape modulations (pumping) of a quantum dot
  in the presence of spin-orbital coupling lead to a finite dc spin
  current. Depending on the strength of the spin-orbit coupling, the
  spin current is polarized perpendicular to the plane of the
  two-dimensional electron gas, or has an arbitrary direction subject
  to mesoscopic fluctuations. We analyze the statistics of the spin
  and charge currents in the adiabatic limit for the full cross-over
  from weak to strong spin-orbit coupling.
\end{abstract}

\maketitle
%%%%%%%%%%
%%%%%%%

There is a growing interest in the physics of spin transport through
low-dimensional quantum structures~\cite{kn:prinz1998} with the aim of
controlling and manipulating spin in microelectronic devices.  A
fascinating tool for spin manipulation is a ``spin battery'' or ``spin
pump'', a device that generates a spin current without an accompanying
charge current~\cite{kn:sharma2001}. Following a proposal of Mucciolo {\em
  et al.}  \cite{kn:mucciolo2002}, a ``spin battery'' was realized recently
\cite{kn:watson2003} using a quantum dot in a two-dimensional electron
gas (2DEG). In this geometry, the direction of spin polarization is
set by an external magnetic field parallel to the plane of the 2DEG.
Current is generated by periodic variation of gate voltages, an
eventual charge current being suppressed by fine tuning the dot shape.
Other proposals for spin pumps have used the idea of locally breaking
spin-rotation symmetry to pump a spin-polarized current by a magnetic
field~\cite{kn:sharma2001}, magnetic
impurities~\cite{kn:aono2002,kn:junling2002,kn:zheng2002}, and by periodic modulation
of the spin-orbit coupling in the 2DEG~\cite{kn:fazio2002}.

Spin current is different from charge current because of the vector
nature of spin. In fact, it is the ``vector'' nature of spin ---
different directions of spin corresponding to different
quantum-mechanical superpositions of ``spin up'' and ``spin down'' ---
that makes it a promising candidate for the practical realization of a
quantum computer~\cite{kn:loss1999}.  Control of magnitude as well as
direction of spin current is of paramount importance if the full
benefits of a spintronic circuit are to be reaped, or if the electron
spin is used as the building block in a scheme that makes essential
use of quantum coherence. The ``spin battery'' of
Refs.~\onlinecite{kn:mucciolo2002,kn:watson2003} satisfies this requirement
in part, since magnitude and sign of the current can be controlled,
but its direction is determined by the external magnetic field,
allowing for spin-polarization in 2DEG plane only.

In this letter, we investigate a quantum-dot based spin pump for which
Bychkov-Rashba~\cite{kn:bychkov1984} and
Dresselhaus~\cite{kn:dresselhaus1955} contributions to spin-orbit
coupling are the sources for spin-rotation symmetry breaking.  We show
that, depending on the strength of the spin-orbit coupling, a quantum
dot with a variable shape can pump either a spin current {\em
  perpendicular} to the 2DEG plane, with the sign and magnitude of the
current subject to control via mesoscopic fluctuations, or a spin
current with an arbitrary direction.

The possibility to generate a spin current perpendicular to the plane
of the 2DEG allows for an interesting realization of a ``spin Hall
effect'' \cite{kn:hirsch1999} if the spin current is injected into a 2DEG
which shows the anomalous Hall effect~\cite{kn:chien1980}. The 2DEG then has
skew spin-orbit scattering with spin up and down (measured
perpendicular to the 2DEG plane) being scattered preferentially in
opposite directions, perpendicular to the flow of current.  While a
current of unpolarized electrons leads to a spin imbalance
perpendicular to the current flow~\cite{kn:hirsch1999} and a spin current
polarized in the plane of the 2DEG is not affected, a spin current
polarized perpendicular to the 2DEG should lead to an anomalous Hall
voltage across the sample.

The system under consideration consists of a ballistic quantum dot
connected to two electron reservoirs through ballistic point contacts
with $N_1$ and $N_2$ channels each, see Fig.\ \ref{var}, inset.
Gate voltages $x_1$ and $x_2$ of two
shape-distorting gates allow for a time-dependent variation of
the dot shape. The same geometry was used in the ``spin battery'' of
Ref.\ \onlinecite{kn:watson2003}, with a quantum dot without notable
spin-orbit coupling. Electron motion in the dot is characterized by the
transit time $\tau_{\rm tr} = L/v_F$, where $L$ is the dot size and
$v_F$ the Fermi velocity, and the time $\tau_{\rm esc}
= h/N\Delta$ for escape to the reservoirs, 
where $N=N_1+N_2$ and $\Delta$ is the mean spacing
between single-electron levels in the quantum dot. The escape time,
which is determined by the point contacts, is typically much larger 
than $\tau_{\rm tr}$. In the absence of spin-orbit coupling, 
periodic variation of the gate voltages $x_1$ and $x_2$ leads to
a dc charge current through the dot
\cite{kn:brouwer1998,kn:zhou1999,kn:switkes1999}. 
Below, we show that a spin
current is generated in the presence of spin-orbit scattering.

The Dresselhaus and Bychkov-Rashba contributions to the spin-orbit coupling
arise from the spin splitting of the conduction band in bulk GaAs 
\cite{kn:pikus1995} and
the asymmetry of the potential well forming the 2DEG, respectively.
The corresponding term in the Hamiltonian has the form
\begin{equation}
  H_{\rm SO} = \varrho\left[
  p_x\sigma_x-p_y\sigma_y\right]
  + \alpha\left[\vec
  p\times\vec\sigma\right]_z,
\end{equation}
where $p_x$ and $p_y$ is the in-plane electron momentum and $\sigma_x$,
$\sigma_y$, and $\sigma_z$ are the Pauli matrices. The
coefficients $\varrho$ and $\alpha$ define a length scale
$$
  \lambda = \frac{\hbar}{m |\varrho^2 - \alpha^2|^{1/2}},
$$ where $m$ is the electron mass. For quantum wires, only the spin
projection perpendicular to the wire and in the plane of the 2DEG is
conserved. As a result, pumping through such a system yields a spin
current with the polarization direction in the plane of the
2DEG~\cite{kn:fazio2002}. For quantum dots of size $L \ll \lambda$, the
role of spin-orbit scattering is qualitatively different from
one-dimensional or bulk systems. The difference originates from the
fact that, although spin-orbit scattering has a small effect in the
transit time $\tau_{\rm tr}$ if $L \ll \lambda$, it still may have a
large effect within the time scale $\tau_{\rm esc}$
\cite{kn:aleiner2001}. As shown in Ref.\ \onlinecite{kn:aleiner2001},
a unitary transformation casts $H$ into the form
\begin{equation}
  H =\frac{1}{2m}(\vec p-\vec a_\perp-\vec a_\parallel)^2,
  \label{eq:H}
\end{equation} 
where 
\begin{eqnarray}
  \vec a_\perp= \frac{\hbar\sigma_z\left[\vec r\times\hat
  z\right]}{2\lambda^2}, \ \
  \vec a_\parallel= \frac{\vec a_\perp}{\hbar}
  \left(\alpha\vec r\cdot\vec\sigma+\varrho[\vec
  r\times\vec\sigma]_z\right)
\end{eqnarray}
 are spin-dependent vector
potentials representing the effects of spin-orbit scattering.
They are characterized by scattering times $\tau_{\perp}$ and
$\tau_{\parallel}$,
\begin{equation}
  \tau_{\perp} = \kappa^{-1} \tau_{\rm tr} (2 \lambda/L)^4,\ \
  \tau_{\parallel} = (\kappa')^{-1} \tau_{\perp} (2 \lambda/L)^2
  \gg \tau_{\perp},
\end{equation}
where $\kappa$ and $\kappa'$ are geometry-dependent coefficients
of order unity \cite{kn:cremers2003}.
Whether or not the spin-orbit
term $a_{\parallel}$ is important depends on the relative size
of $\tau_{\parallel}$ and $\tau_{\rm esc}$. We now discuss the 
cases $\tau_{\parallel} \gg \tau_{\rm esc}$
and $\tau_{\parallel} \lesssim \tau_{\rm esc}$ separately.

For $\tau_{\parallel} \gg \tau_{\rm esc}$ the only relevant spin-orbit
term in the Hamiltonian is the spin-dependent vector potential
$a_{\perp}$. This term has the same form as the vector potential
arising from a magnetic field of size $B_{\rm so} = \hbar/2 e
\lambda^2$ with opposite directions for spin up ($\up$) and spin down
($\dwn$), measured perpendicular to the plane of the 2DEG. Hence, the
perpendicular spin projections are conserved, and separate dc currents
$I_{\uparrow}$ and $I_{\downarrow}$ will be pumped through the quantum
dot for spin up and spin down. Generically, one has a spin current
$I_{{\rm s},z} = I_{\uparrow} - I_{\downarrow} \neq 0$ because
electron trajectories that enclose the same geometrical area (see
inset of Fig.~\ref{var}) gather an ``Aharonov-Bohm'' flux with opposite
signs for electrons with opposite spin projections, thereby giving the
scattering phase shift a spin dependence. In a chaotic cavity, both
the sign and magnitude of the currents $I_{\up,\dwn}$ is essentially
random, dependent on the detailed shape and electron density of the
quantum dot. Following the protocol of Ref.~\onlinecite{kn:mucciolo2002}, a
spin battery with spin current perpendicular to the 2DEG plane results
if the shape of the dot is fine tuned so that the charge current
$I_{\rm c} = I_{\uparrow} + I_{\downarrow} = 0$.  In order to compare
typical (rms) sizes of spin and charge currents, we note that the spin
current is formally equal to the magnetic-field antisymmetric
component of the charge current for a quantum pump with spinless
electrons in a perpendicular magnetic field of size $B_{\rm so}$. The
latter quantity was calculated in Ref.\ \cite{kn:shutenko2000} using
random matrix theory, which is valid if the dot has an irregular shape
and $\tau_{\rm esc} \gg \tau_{\rm tr}$,
\begin{equation}
 {\rm rms}\, I_{{\rm s},z}
  = \left(1 - \frac{1}{(1 + \tau_{\rm esc}/\tau_{\perp})^3} 
  \right)^{1/2} {\rm rms}\, I_{\rm c}.
\end{equation}

We now turn to the general case $\tau_{\parallel} \lesssim
\tau_{\rm esc}$. We consider a sinusoidal time dependence of
the gate voltages, $x_1(t) = \delta x_1 \cos(\omega t)$ and $x_2(t) =
\delta x_2 \cos(\omega t + \phi)$ and consider the contribution
to the charge and spin current that is bilinear in the gate voltage
amplitudes $\delta x_1$ and $\delta x_2$. Starting point of our
calculation is the relation between the charge and spin currents
and the $2N\times 2N$ scattering matrix $S$ of the quantum dot
\cite{kn:brouwer1998},
\begin{eqnarray}
  I_{\rm c} &=& {\omega  \delta x_1 \delta x_2 \sin \phi\over 2\pi}\,
  {\rm tr}\,\mbox{Im}\left[(\Lambda \otimes \mathbb{1})
  \;{\partial S\over\partial x_2}
  \;{\partial S^{\dg}\over\partial x_1}\right], \nonumber \\
  I_{\rm s} &=& {\omega \delta x_1 \delta x_2 \sin \phi \over 2\pi}\,
  {\rm tr}\,\mbox{Im}\left[(\Lambda \otimes \vec \sigma)
  \;{\partial S \over \partial x_2}
  \;{\partial S^{\dg}\over\partial x_1}\right],
  \label{eq:current}
\end{eqnarray}
where $\Lambda$ is an $N \times N$ diagonal matrix with elements
$\Lambda_{jj}=N_2/N$ for $j\leq N_1$, and $\Lambda_{jj}=-N_1/N$ 
for $N_1<j\leq N$ and $\openone$ is the $2 \times 2$ unit matrix
in spin space. 

In order to calculate the average and root-mean-square (rms) of the
charge and spin current for an ensemble of quantum dots, we need to
know the average of a product of up to four scattering matrix elements
taken at different values of the parameters $x_1$ and $x_2$.  We
calculate this correlator using random matrix theory (RMT)
\cite{kn:beenakker1997}. In RMT, the spin-orbit part of the
Hamiltonian (\ref{eq:H}) is replaced by a $2 M \times 2 M$ random
hermitian matrix \cite{kn:cremers2003}
\begin{eqnarray}
  \label{eq:calh}
  H_{\rm so}
    &=& i\sqrt{\frac{\Delta}{4 \pi}}
  \left[\frac{A_3 \otimes \sigma_3}{\sqrt{\tau_{\perp}}}
  + \frac{A_1 \otimes \sigma_1 + A_2 \otimes \sigma_2}
  {\sqrt{\tau_{\parallel}}}\right],
\end{eqnarray}
whereas the dependence on the shape-distorting gate voltages $x_1$
and $x_2$ is represented through the random hermitian matrix
\begin{eqnarray}
  \label{eq:calhx}
  H_{\rm shape} &=& {\Delta\over\pi} \sum_{j=1}^{2} x_j X_j \otimes
  \mathbb{1}.
\end{eqnarray}
Here, $X_j$, $j=1,2$ are real symmetric random $M \times M$
matrices and 
$A_{j}$, $j=1,2,3$ are real antisymmetric random $M \times M$
matrices with 
\begin{equation}
  {\rm tr}\, X_i X_j = {\rm tr}\, A_i A_j = M^2
  \delta_{ij}. \label{eq:trdef}
\end{equation} 
At the end of the calculation, the limit $M \to\infty$ needs to be
taken. Performing the random matrix average using standard methods
\cite{kn:polianski2003,kn:cremers2003}, we find that the relevant
average of a product of four scattering matrices is
\begin{eqnarray}
 \lefteqn{\la S(1)_{kl} \otimes S^{\dg}(2')_{nk} \otimes
  S(1')_{mn} \otimes S^{\dg}(2)_{lm} \ra}
  \nonumber \\ &\equiv& W^{(1)}(12') W^{(1)}(1'2) \delta_{ln}
  + W^{(2)}(12'1'2), \label{eq:SSSS}
  ~~~~~
\end{eqnarray}
where ``1'', ``2'', ``$1'$'', and ``$2'$'' are shorthand notations
for values of the gate voltages $x_1$ and $x_2$ and the Fermi
energy $\varepsilon$ and we have used tensor notation for the
spin degrees of freedom. The first contribution on the r.h.s.\
of Eq.\ (\ref{eq:SSSS}) is the product of pair averages 
\cite{kn:cremers2003},
\begin{eqnarray}
  W^{(1)}(12')
  &=& \frac{1}{M \openone_2 - \mbox{tr}\,
  R(1) \otimes R^{\dagger}(2')}
  \otimes \openone_2 \nonumber,
  \label{eq:W1}
\end{eqnarray}
with a similar definition of $W^{(1)}(1'2)$. Here $\openone_2\equiv
\openone\otimes\openone$, the auxiliary matrix $R$ is defined as
\begin{equation}
  R(\varepsilon,x_1,x_2) = 
  \exp\left[ 2 \pi i(\varepsilon - H_{\rm so} - H_{\rm shape}(x_1,x_2))
  \right], \label{eq:R}
\end{equation}
and the trace ``tr'' is not taken over the spin degrees of freedom.
In taking the inverse in Eq.\ (\ref{eq:W1}), the rule for
multiplication of the tensor products is $(\sigma_j\otimes\sigma_{k'})
(\sigma_{k}\otimes\sigma_{j'}) = (\sigma_j\sigma_{k})\otimes
(\sigma_{j'}\sigma_{k'})$. The second contribution in Eq.\
(\ref{eq:SSSS}) involves the random matrix equivalent of the
Hikami box from diagrammatic perturbation theory and can be
calculated using the method of Ref.\ \onlinecite{kn:polianski2003}.
Using the tensor notation with the same product rules as before,
the result can be written as
\begin{eqnarray}
  W^{(2)}(12'1'2)&=& W^{(1)}(12') W^{(1)}(1'2)
  {\cal D}(12'1'2)
  \nonumber \\ && \mbox{} \times
  W^{(1)}(12) W^{(1)}(1'2'),
\end{eqnarray}
where
\begin{eqnarray*}
  {\cal D}(12'1'2)=M\mathbb{1}_4 -{\rm tr}\;R(1)\otimes R^{\dg}(2')\otimes R(1')\otimes R^{\dg}(2),
\end{eqnarray*}
The traces are calculated using Eq.\ (\ref{eq:trdef}) after expanding
$R$ to second order in $H_{\rm so}$ and $H_{\rm shape}$ and taking $M
\to \infty$.

Substituting these results into Eq.\ (\ref{eq:current}) one then finds
the average and variances of the spin and charge currents. All
ensemble averages are zero, as well as the covariances for different
components of the spin current or of spin and charge currents. The
full results for the average square spin and charge currents can be
written by introducing $I_0 = (\omega/2 \pi)4\delta x_1 \delta x_2
\sin \phi \sqrt{N_1 N_2/N^4}$, $c_{\perp} = \tau_{\rm
  esc}/\tau_{\perp}$, and $c_{\parallel} = \tau_{\rm
  esc}/\tau_{\parallel}$:
\begin{eqnarray}
\langle I_{\rm c}^2 \rangle
  &=&
  I_0^2 
  \left(1 +\frac{1}{(1+2c_\parallel)^3}+\frac{2}{(1+c_\parallel+
      c_\perp)^3}\right), \nonumber \\
  \langle I_{{\rm s},z}^2 \rangle
  &=& I_0^2 
  \left(1 +\frac{1}{(1+2c_\parallel)^3}-\frac{2}{(1+c_\parallel+
      c_\perp)^3}\right), \nonumber \\
  \langle I_{{\rm s},x}^2 \rangle
  &=&\langle I_{{\rm s},y}^2 \rangle=
  I_0^2 
  \left(1 - \frac{1}{(1 + 2c_{\parallel})^3}
  \right)
\end{eqnarray}

These general results confirm that spin current is polarized
perpendicular to the plane of the 2DEG if $\tau_{\parallel} \gg
\tau_{\rm esc}$, while its direction is arbitrary if
$\tau_{\parallel} \ll \tau_{\rm esc}$. In practice, the
ratio $\tau_{\parallel}/\tau_{\rm esc}$ can be tuned rather
straightforwardly by changing the conductances of the point
contacts connecting the quantum dot to the outside world or
the dot size. Recent experiments by Zumb\"uhl {\em et al.}
show that both limits can be obtained experimentally
\cite{kn:zumbuhl2002}: Depending on the size of the quantum dots,
Zumb\"uhl {\em et al.} find $c_{\perp} N$ ranging from $0.94$ to
$20$ and $c_{\parallel} N$ ranging from $0.25$ to $20$. 
For $N=2$ and at a temperature
$T\ll \tau_{\rm esc}^{-1}$, we obtain rms $I_{{\rm s},z}$/rms $I_{\rm c}
= 0.72$. At finite temperature, both the
effect of thermal smearing and a finite dephasing time $\tau_{\phi}$
need to be taken into account. Dephasing is accounted for by the
substitution
$1/\tau_{\rm esc} \to 1/\tau_{\rm esc} + 1/\tau_{\phi}$, whereas
thermal smearing requires integration over $\varepsilon$. For large
temperatures $T\gg \tau_{\rm esc}^{-1}$, and in the absence of any Zeeman
coupling, we can borrow the results of Ref.~\cite{kn:shutenko2000} and
find the polarization ratio rms $I_{{\rm s},z}/{\rm rms} I_{\rm c} = 
[(c_{\perp}^2 + 2 c_{\perp})/(c_{\perp}^2 + 2 c_{\perp} + 2)]^{1/2}
= 0.61$.

Before concluding, we discuss the dependence of the pumped current on
an applied magnetic field $\vec B$.  This question is relevant if a
spin pump is used in conjunction with spin manipulation by means of
in-plane or perpendicular fields. Within RMT, the effect of a magnetic
field is modeled by a third random matrix
\cite{kn:aleiner2001,kn:cremers2003}
\begin{equation}
  H_{B} = \sqrt{\frac{\Delta}{4 \pi}} 
  \left( i \frac{A_0 \otimes \openone}{\sqrt{\tau_{H}}}
  + \frac{X \otimes \sigma_z}{\sqrt{\tau_{H,\perp}}}
  \right) - \frac{1}{2\tau_Z} \hat B \cdot \vec \sigma,
\end{equation}
where $1/\tau_Z=\mu_B g B$, $B$ is the magnitude of the applied
magnetic field, $\hat B$ its direction, $\mu_B$ is the Bohr magneton,
$g$ is the $g$ factor of GaAs, $\tau_H = \kappa^{-1} \tau_{\rm tr}
(2/e B_z L^2)^2$, $\tau_{H,\perp} = (4 \lambda/\mu_B g L
B)^{2}/\kappa'' \tau_{\rm tr}$, $\kappa$ and $\kappa''$ being
geometry-dependent coefficients of order unity, $A$ is an $M \times M$
real antisymmetric matrix with trace $\mbox{tr}\, A A^{\rm T} = M^2$,
and $X$ is a real symmetric matric with trace $\mbox{tr}\, X^2 = M^2$.
The time $\tau_{H}$ describes the orbital effect of the magnetic field
component perpendicular to the 2DEG; it is the time needed for picking
up a quantum of magnetic flux.  The time $\tau_{H,\perp}$ describes
spin-flip processes arising from the interplay of the component of the
magnetic field parallel to the 2DEG and the spin-orbit scattering
\cite{kn:halperin2001}.

Adding the random matrix $H_{B}$ to the exponent in Eq.\ (\ref{eq:R})
and repeating the previous calculations, we find (i) inclusion of
$\tau_{H}$ has no effect on the current statistics and (ii) inclusion
of $\tau_{H,\perp}$ amounts to the subsitution $1/\tau_{\perp} \to
1/\tau_{\perp} + 1/\tau_{H,\perp}$. In general the effect of Zeeman
coupling to the parallel field (time scale $\tau_Z$) is to decrease
correlations between $I_{\up}$ and $I_{\dwn}$, so that both rms
$I_{{\rm s},z}$ and rms $I_{c}$ are reduced. On the other hand, the
variance of the spin current polarized along the direction of the
in-plane magnetic field is enhanced. Both these effects are shown in
Fig.~\ref{var}.  The full results for the current variance in the
presence of Zeeman coupling are rather lengthy and will be reported
elsewhere.  Here, we confine ourselves to the limiting cases of
Ref.~\cite{kn:aleiner2001} which are distinguished by a parameter
$\Sigma=1,2$ characterizing the mixing of states with different spins
for strong Zeeman splitting. Choosing the magnetic field along the $x$
direction, these are: (i) $c_\parallel=\tau_{\rm
esc}/\tau_\parallel\ll 1$, and $b\equiv\tau_{\rm esc}/\tau_Z$ large or
small in comparison to $c_\perp=\tau_{\rm esc}/\tau_\perp$, for which
$\Sigma=1$; and (ii) $c_\perp\gg 1$, $b$ large or small in comparison
to $c_\parallel$, for which $\Sigma=2$.  For the case (i), when
$b^2/c_\perp\ll 1\ll c_\perp$, we obtain to the lowest order:
\begin{figure}
  \includegraphics[scale=.32]{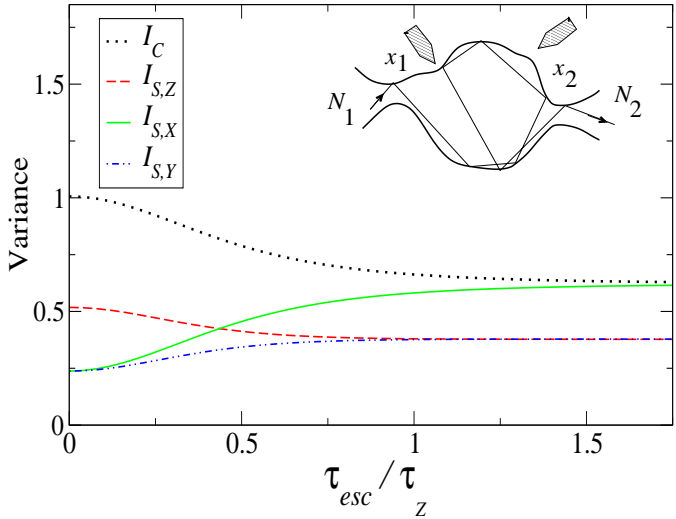}
  \caption{Current variance (in units of $I_0^2/2$) as a function of
    Zeeman coupling strength ($b=\tau_{\rm esc}/\tau_Z$) for a fixed
    value of SO coupling strengths $c_{\perp}=0.47$, and
    $c_\parallel=0.12$ taken from Ref.~\cite{kn:zumbuhl2002}. The
    difference between $\la I_{{\rm s},x}^2\ra$ and $\la I_{{\rm
        s},z}^2\ra$ for large $b$ vanishes for $c_\perp\gg 1$. Inset
    shows a quantum dot with out of phase shape modulation, and a
    typical electron trajectory which gathers a Aharonov-Bohm flux
    depending on its perpendicular spin projection.}
\label{var}
\end{figure}
\begin{eqnarray}
\label{smallb}
  \la I_{c}^2 \rangle &=&\la I_{{\rm s},z}^2\ra =   I^2_0 \left(2 
  - 12 b^2/c_\perp + 96 b^4/c_\perp^2\right) ,\nonumber \\
  \la I_{{\rm s},x}^2\ra&=&\langle I_{{\rm s},y}^2 \rangle
  = I^2_0\left( 12 b^2/c_\perp - 96 b^4/c_\perp^2\right),
  \label{smallb-x}
\end{eqnarray}
and when $c_\perp\ll 1 \ll b$ we find:
\begin{eqnarray}
  \la I_{c}^2 \ra &=&\la I_{{\rm s},x}^2 \ra=
  I_0^2 \left( 2-3c_\perp + 6c_\perp^2\right)
  \nonumber \\
  \la I_{{\rm s},z}^2 \ra &=&\la I_{{\rm s},y}^2 \ra =
  I_0^2 \left( 3c_\perp-6c_\perp^2 \right),
\end{eqnarray}
which is ``dual'' to expression (\ref{smallb}) under the exchange
$I_{{\rm s},x}\leftrightarrow I_{{\rm s},z}$ and $c_\perp\rightarrow
4b^2/c_\perp$. For the case (ii), the lowest order expansion in
$b^2/c_\perp\ll 1\ll c_\perp,c_\parallel$:
\begin{eqnarray}
  \la I_{c({\rm s},x)}^2 \ra &=&
 I_0^2 \left( 1\pm 3/8\;c_\perp^{-3}\mp 1/2 (b^2/c_\perp) c_\perp^{-4} \right)
  \nonumber \\
  \la I_{{\rm s},z}^2 \ra &=&\la I_{{\rm s},y}^2 \ra =
  I_0^2 \;\left(1-1/8\;c^{-3}_\perp\right),
\end{eqnarray}
is dual under the transformation $4b^2/c_\perp\to c_\parallel$ to the
variance in the other limit $c_\parallel\ll 1\ll c_\perp,b^2/c_\perp$.
The non-vanishing current variances $\la I^2\ra$ can be written in a
unified way as $\la I^2\ra=4(s/\beta\Sigma)I_0^2$, where $\beta=2$
defines the time reversal symmetry of orbital motion, and $s=1$ is the
Kramers degeneracy parameter.

In conclusion, we have shown that a quantum pump consisting of a
quantum dot with spin-orbit coupling allows for the generation of both
a dc charge current and a dc spin current. Depending on the strength
of the spin-orbit scattering or on the conductances of the point
contacts between the dot and the reservoirs, the pumped spin current
is perpendicular to the 2DEG or has an arbitrary direction.
Measurement of the spin current is possible via the spin Hall effect
or by connection of the spin pump to a (semiconducting) ferromagnet
with known magnetization direction.  Furthermore, the method of
adiabatic spin pumping, as opposed to other mechanisms such as
rectification~\cite{kn:brouwer2001,kn:watson2003}, allows the direction of
spin polarization to be continuously changed.
\begin{acknowledgements}
  We thank Charles Marcus, Claudio Chamon, Karsten Flensberg and
  Vladimir Fal'ko for discussions.  This work was supported by the NSF
  under grant no.~DMR 0086509, and by the Packard Foundation.
\end{acknowledgements}
%
%%%%%%%%%%%%%%%%%%%%%%%%%
% REFERENCES
%%%%%%

%%%%%%%%%%%%%%%%%%%%%%%%%%
%%%%%%%%%%%%%%%%%%%%%%%%%%
\end{document}